%% file: 000_main.tex
\documentclass[sigconf,review=false, balance=false, anonymous=false]{acmart}
\usepackage{popets}

\input{macro}

\usepackage{tabularx,booktabs}

\newcommand\blfootnote[1]{%
  \begingroup
  \renewcommand\thefootnote{}\footnote{#1}%
  \addtocounter{footnote}{-1}%
  \endgroup
}

\usepackage{fontawesome}
\makeatletter
\def\DeclareUnicodeCharacter#1#2{}
\makeatother

\usepackage{utfsym}

\usepackage{ragged2e}

\usepackage{etoolbox}
\makeatletter
\patchcmd{\@makecaption}
  {\scshape}
  {}
  {}
  {}
\makeatother

\usepackage{mwe}

\usepackage{array}

\usepackage{multirow}
\usepackage{makecell}

\usepackage{pifont}

\usepackage{amsmath}
\usepackage{bm}
\usepackage{amssymb}
\usepackage{amsthm}

\makeatletter
\newcommand{\removelatexerror}{\let\@latex@error\@gobble}
\makeatother

\usepackage{comment}  

\usepackage{multicol,lipsum,environ}

\NewEnviron{auxmulticols}[1]{%
  \ifnum#1<2\relax
    \BODY
  \else
    \begin{multicols}{#1}
      \BODY
    \end{multicols}%
  \fi
}

\definecolor{lavender}{rgb}{0.75, 0.58, 0.89}
\newif\ifshowcomment
\showcommenttrue
\ifshowcomment
    \newcommand{\yujin}[1]{\textsf{\color{blue}{[{Yujin: #1}]}}}
    \newcommand{\chris}[1]{\textsf{\color{magenta}{[{Chris: #1}]}}} \newcommand{\ella}[1]{\textsf{\color{red}{[{Ella: #1}]}}}
    \newcommand{\natacha}[1]{\textsf{\color{magenta}{[{Natacha: #1}]}}}
\else
    \newcommand{\dawn}[1]{}
    \newcommand{\yujin}[1]{}
    \newcommand{\chris}[1]{}
    \newcommand{\ella}[1]{}
    \newcommand{\natacha}[1]{}
    \newcommand{\new}[1]{}
\fi

\usepackage{multicol,lipsum,environ}

\usepackage{amsmath,amsfonts}
\usepackage{graphicx}
\usepackage{textcomp, array}
\usepackage{makecell}
\usepackage{xcolor} 
\usepackage{multirow}
\usepackage{longtable}
\usepackage[T1]{fontenc}
\usepackage[utf8]{inputenc}

\usepackage{balance}
\usepackage{tikz}

\setcopyright{popets}
\copyrightyear{2025}

\acmYear{2025}
\acmVolume{YYYY}
\acmNumber{X}
\acmDOI{XXXXXXX.XXXXXXX}
\acmISBN{}
\acmConference{}
\begin{document}
\title[The Gap Between Data Rights Ideals and Reality]{The Gap Between Data Rights Ideals and Reality}


\author{Yujin Potter}
\affiliation{%
  \institution{UC Berkeley}
  \city{Berkeley}
  \country{USA}}
\email{yujinyujin9393@berkeley.edu}

\author{Ella Corren}
\affiliation{%
  \institution{UC Berkeley, School of Law}
  \city{Berkeley}
  \country{USA}}
\email{ecorren@berkeley.edu}

\author{Gonzalo Munilla Garrido}
\affiliation{%
  \institution{TUM}
  \city{Munich}
  \country{Germany}}
\email{gonzalo.munilla-garrido@tum.de}

\author{Chris Hoofnagle}
\affiliation{%
  \institution{UC Berkeley, School of Law}
  \city{Berkeley}
  \country{USA}}
\email{choofnagle@berkeley.edu}

\author{Dawn Song}
\affiliation{%
  \institution{UC Berkeley}
  \city{Berkeley}
  \country{USA}}
\email{dawnsong@cs.berkeley.edu}


\renewcommand{\shortauthors}{Potter et al.}

\begin{abstract}
As information economies burgeon, they unlock innovation and economic wealth while posing novel threats to civil liberties and altering power dynamics between individuals, companies, and governments. Legislatures have reacted with privacy laws designed to empower individuals over their data. These laws typically create rights for ``data subjects'' (individuals) to make requests of data collectors (companies and governments). The European Union General Data Protection Regulation (GDPR) exemplifies this, granting extensive data rights to data subjects, a model embraced globally.
However, the question remains: do these rights-based privacy laws effectively empower individuals over their data? This paper scrutinizes these approaches by reviewing 201 interdisciplinary empirical studies, news articles, and blog posts. We pinpoint 15 key questions concerning the efficacy of rights allocations. The literature often presents conflicting results regarding the effectiveness of rights-based frameworks, but it generally emphasizes their limitations. We offer recommendations to policymakers and Computer Science (CS) groups committed to these frameworks, and suggest alternative privacy regulation approaches.
\end{abstract}


\maketitle


\blfootnote{\textbf{Disclaimer:} The information and views set out in this article are those of the authors and do not necessarily reflect the official opinion of the institutions. }

\input{001_Intro}
\input{003_Criteria}

\input{004_User}
\input{005_Company}
\input{006_Regulator}
\input{007_Future}
\input{008_Alternative}

\input{AI_privacy}
\input{009_Conclusion}

\newpage
\bibliographystyle{ACM-Reference-Format}
\bibliography{999_References}

\appendix
\input{100_Appendix}

\end{document}

%% file: macro.tex
\usepackage{amsmath,amssymb,amsfonts}
\usepackage{graphicx}
\usepackage{textcomp}
\usepackage{bmpsize}
\usepackage{comment}
\usepackage{enumitem}
\usepackage{amssymb}
\usepackage{pifont}
\usepackage{makecell}

\usepackage{tikz}
 \UseRawInputEncoding

\usepackage{xcolor}
\newcommand{\dawn}[1]{\textcolor{blue}{*Dawn S.: #1}}

\setcitestyle{numbers}

\usepackage{amsthm}
\makeatletter
\def\els@aparagraph[#1]#2{\elsparagraph[#1]{#2\@addpunct{.}}}
\def\els@bparagraph#1{\elsparagraph*{#1\@addpunct{.}}}
\makeatother

\usepackage{caption} 
\usepackage{float}
\floatstyle{plaintop}
\restylefloat{table}

\usepackage{xcolor}

\usepackage{array}
\newcolumntype{M}[1]{>{\centering\arraybackslash}m{#1}}

\usepackage{afterpage}

\usepackage{comment}

\usepackage{booktabs}

\usepackage{pdfpages}

\usepackage{algorithmicx}
\usepackage[ruled, vlined]{algorithm2e}
\usepackage{float}

\usepackage{scalerel,stackengine}
\stackMath
\newcommand\reallywidehat[1]{%
\savestack{\tmpbox}{\stretchto{%
  \scaleto{%
    \scalerel*[\widthof{\ensuremath{#1}}]{\kern-.6pt\bigwedge\kern-.6pt}%
    {\rule[-\textheight/2]{1ex}{\textheight}}
  }{\textheight}%
}{0.5ex}}%
\stackon[1pt]{#1}{\tmpbox}%
}

\usepackage{color}

%% file: 001_Intro.tex
\section{Introduction}
\label{sec:intro}

The digital age introduces significant privacy threats and challenges. Policymakers have largely responded with rights-based regimes that grant users specific rights, such as accessing and erasing their data, aiming to empower them in the face of surveillance capitalism.

These regimes, now a regulatory standard enacted in numerous global privacy laws~\cite{solove2023rights}, primarily follow the European Union General Data Protection Regulation (GDPR) model, which grants the most extensive set of data rights~\cite{solove2023rights,hoofnagle2019european} (see Appendix Tab.~\ref{tab:right}). The U.S. has adopted similar approaches, as seen in the California Consumer Privacy Protection Act (CCPA), modeled after the GDPR.

The proposition that bestowing individuals with data rights can mitigate corporate malfeasance and safeguard personal privacy is a subject of disagreement~\cite{solove2023rights, waldman2021new, lazaro2015control}. Many hold a positive position on rights regimes, but the traditional belief hinges on an intricate, contingent network of assumptions regarding user awareness, the capacity to exercise their rights, and the companies' diligence in offering, elucidating, and applying controls. 
\newline
However, rights are not self-actuating; individuals need to be cognizant of their rights and possess the capability to exercise them. 
In addition, in instances where infringing upon rights can provide advantages to companies, these corporations frequently establish transaction costs and other obstacles to dissuade or reject user rights invocations~\cite{janger2001gramm}. Furthermore, the implementation of rights-based approaches needs enforcement actions, which can be subject to either insufficient or excessive enforcement by individuals and regulators.
\newline 
Such opposing viewpoints on the effectiveness of rights-based regimes call for a more comprehensive examination of the field. Against this backdrop, this paper presents a meta-analysis of existing empirical evidence, evaluating the effectiveness and desirability of these regimes. Our analysis uncovers a significant variability in the efficacy of data rights, contingent upon the user, company, regulator, and context. While acknowledging the beneficial role data rights have played in certain situations, we argue that their overall effectiveness remains limited.
\newline 
To provide a comprehensive perspective on the efficacy and desirability of individual privacy rights, we develop an evaluation framework synthesizing 201 academic papers, news articles, and blog posts. 
Our investigation raises 15 key questions, revealing conflicting narratives and indicating that the current implementation of rights-based regimes needs improvement. 
We conclude by recommending potential enhancements and exploring alternatives to rights-based regimes.

%% file: 003_Criteria.tex
\section{Data Rights Assessment Framework \& Method}
\label{sec:method}

Since the European data rights model has become the standard in the field~\cite{solove2023rights, waldman2021new, hoofnagle2019european}, we select the GDPR data rights in Tab.~\ref{tab:right} as the focus of our analysis.
We developed a general data rights assessment framework consisting of 15 key questions (see Appendix \ref{app_sec:Framework_questions}) by synthesizing 201 interdisciplinary empirical materials. This framework provides a comprehensive evaluation of rights-based privacy approaches, considering the perspectives of the primary actors in the information economy: individuals, companies or developers, and regulators. Practitioners and researchers can use our three pronged framework as a starting point to analyze other data-rights frameworks like CCPA.

Based on the extant empirical studies, we discuss the 15 questions for the three actors (users, companies, and regulators) and reveal open challenges in \S\ref{sec:user}, \S\ref{sec:company}, and \S\ref{sec:regulator}. For each question, we examine whether there is divergent evidence and, if so, \textit{why}. Tab.~\ref{tab:framework} summarizes the evaluation results. 

\smallskip\noindent\textbf{Method:} We initiated our research with a preliminary review of 10 papers pertaining to GDPR data rights, already familiar to the authors. Guided by these papers, the authors constructed the search keywords \texttt{\small "GDPR" AND "privacy” AND ("right*" OR "control*" OR "setting*" OR "polic*" OR "dashboard*")}, designed to capture papers containing phrases such as GDPR data rights, control over data (or data control), privacy policy (referring to the right to information), privacy settings/data dashboards (representing terms relevant to data management), and so forth.
\newline 
To ascertain a sufficiently diverse selection of papers across disciplines, including computer science and social sciences, we collated papers from three databases: ScienceDirect, Scopus, and the ACM library, using the aforementioned keywords. This search yielded an initial collection of 2322 articles, with 68, 988, and 1266 articles from ScienceDirect, Scopus, and the ACM library, respectively.

After eliminating duplicate entries, we were left with a primary dataset of 2234 articles. We subsequently filtered this dataset through a three-step process: scrutinizing the title, abstract, and finally the full text. Our focus was primarily on papers presenting empirical data directly pertinent to the eight GDPR data rights, rather than those broadly related to general GDPR policy or privacy topics.
Consequently, we examined papers for empirical evidence concerning GDPR data rights and aligned each piece of evidence with the corresponding data rights. Our analysis also incorporated studies of the eight data rights implemented outside of Europe. The rigorous application of these selection criteria culminated in a final collection of 150 papers out of the initial 2234.


Beyond the database search, we also examined the references cited in some of the identified papers and, as a result, incorporated additional relevant articles. 
Leveraging the authors' domain expertise, we integrated further articles and related works, including public opinion surveys or news articles that encompass extensive empirical data or support our gathered empirical evidence. Moreover, we studied the websites of Data Protection Authorities (DPAs) in six European countries - France, Germany, Norway, Portugal, Spain, and the UK - to examine regulators' activities concerning GDPR data rights. Through the process, we incorporated additionally 20 academic papers and 31 non-academic public sources. 
As a result, our final dataset comprises 201 resources, inclusive of 170 academic papers and 31 other public materials (e.g., public reports, news articles, blog posts, and DPA websites), all of which focus on GDPR data rights and contain empirical data. 

Utilizing such data, we executed a reflexive thematic analysis~\cite{braun2019reflecting} and iteratively developed a codebook. The completed codebook is displayed in Tab.~\ref{tab:code}. 
The coding process and categorization of the papers were initially executed by one Computer Science (CS) researcher. Subsequently, two CS researchers, drawing from the resulting codes and categories, formulated the 15 core questions constituting our data rights evaluation framework. Finally, through iterative discussions, two Law researchers refined and enhanced the framework. 
Through the lens of these 15 questions, we assessed the current state and effectiveness of rights-based regimes and identified positive and dysfunctional components within these regimes.

%% file: 004_User.tex
\section{Users' perspective}
\label{sec:user}

First, we explore the four key questions related to users' understanding, perception, exercise, and benefits of their data rights.

\smallskip
\noindent \textbf{(U1) Are users informed about and understand data rights?}

\indent (U1.1) \textit{Disparity in data rights knowledge}. 
The understanding of data rights is not uniform, with some rights being more recognized than others~\cite{eurobarometer,rughinis2021social,kuebler2021right}. A large-scale user survey by Eurobarometer~\cite{eurobarometer} reveals that the least known rights are the right to data portability and the right to avoid automated decision making. However, the rights to access, erasure, and object are relatively well-known. Consistent results emerge from a German study~\cite{kuebler2021right}.
\smallskip

\indent (U1.2) \textit{The digital divide and its impact}.
There is a correlation between internet access and GDPR knowledge \cite{eurobarometer}. Non-internet-users, or ``offline citizens'', display minimal awareness, with only 7.2\% cognizant of the GDPR. In contrast, ``data citizens'' who frequently use the Internet exhibit higher GDPR knowledge~\cite{rughinis2021social, eurobarometer}. 
There is a geographic trend, too, with countries having higher internet penetration reporting more GDPR-aware citizens~\cite{sideri2020we}.
Rughini\textcommabelow{s} et al.~\cite{rughinis2021social} found that the most rights-aware Europeans live in The Netherlands, Ireland, and the UK, while the low-awareness users are clustered in low-internet-penetration nations, Bulgaria, Romania, and Greece.
Exploring the Greek landscape~\cite{sideri2020we}, we observe that internet usage promotes understanding of data rights more effectively than traditional mass media or word-of-mouth. This scenario implies that the digital divide contributes to uneven GDPR knowledge.
\smallskip 

\indent (U1.3) \textit{Contextual variation in knowledge}. 
The awareness level also changes dramatically depending on the technology context~\cite{lau2018alexa,reis2018patients,liao2020measuring,sun2021they}.
For instance, many users understand how to exercise the right to access with smart speakers like Amazon Echo and Google Home~\cite{lau2018alexa}, yet only 24\% of patients are aware of accessing their health information~\cite{reis2018patients}. Moreover, many users are uninformed about the availability of privacy policies of their voice apps~\cite{liao2020measuring}. Recall the fact that the right to information and access are well-known in general~\cite{pew-privacy, eurobarometer,rughinis2021social,kuebler2021right}. As a result, this discrepancy highlights the gap between abstract knowledge and practical application of data rights.
\smallskip

\indent (U1.4) \textit{Impact of education}. 
Educational interventions appear to foster data rights awareness. An experiment demonstrates that a simple informative video can significantly enhance the understanding of the right to avoid automated decision making~\cite{kaushik2021know}: 80\% of US participants and 64\% of UK participants' understanding became ``extremely clear''. Likewise, a study among university students reports moderate to high awareness of data rights, potentially due to exposure to related courses in that university~\cite{presthus2018consumers}.
\smallskip

%



\noindent \textbf{(U2) How do users perceive data rights?}
\smallskip

\indent (U2.1) \textit{Positive perception}. 
Across numerous contexts, we find that users often hold a positive perception and express interest in their data rights, ranging from the right to information to the right to avoid automated decision making~\cite{soonthornphisaj2019internet, presthus2018consumers, lau2018alexa, reis2018patients, owens2021you, bruggemeier2022perceptions, alizadeh2020gdpr, presthus2019consumer, kaushik2021know, harambam2019designing, kulyk2020has, jennes2017social, eiband2018bringing, zhang2021facial, zhang2021did}.
For instance, users appreciate being informed about their data rights by service providers~\cite{bruggemeier2022perceptions, zhang2021facial, zhang2021did}. They often express curiosity about the reasons for data collection and how this data is utilized by firms~\cite{lau2018alexa, harambam2019designing, kulyk2020has, nunes2020futures, bowyer2018understanding, marky2020all, dowthwaite2020s}. Indeed, depending on the context, users want different types of information and different notification frequencies~\cite{ebert2020does, zhang2021did}. 
In an interview regarding a retail loyalty card program, some consumers considered it beneficial to access the personal information collected about them~\cite{alizadeh2020gdpr}. Similarly, patients acknowledged the value of the right to access and expressed a desire to exercise this right~\cite{reis2018patients}. Hence, many users perceive the right to information and access positively and wish to exercise these rights.
\newline 
The desire to exercise the right to erasure is also prevalent~\cite{alizadeh2020gdpr, nunes2020futures, bowyer2018understanding, sun2021child}. 
For example, despite being under a legal surveillance regime, some surveilled subjects (or their families) felt they should have the right to delete their data~\cite{owens2021you}. Approximately 89\% of users in another study wanted the ability to delete their data and verify the erasure~\cite{mangini2020empirical}.
A favorable view is also held for the right to object. Some survey participants appreciated the ability to opt-out of receiving direct marketing~\cite{presthus2019consumer}. Moreover, users acknowledged the importance of the right to avoid automated decision-making~\cite {presthus2018consumers,presthus2019consumer, kaushik2021know, sankaran2021s, vaccaro2020end}. For example, some users preferred making independent choices, even if they were merely suggestions or recommendations on a website~\cite{sankaran2021s}. The desire to avoid automated decision-making heightened when it involved personal matters like insurance, mortgage loans, or exam results evaluation~\cite{presthus2018consumers, presthus2019consumer, kaushik2021know, vaccaro2020end}.
\smallskip

\indent (U2.2) \textit{Negative perception}. 
However, not all users desire data rights. Some individuals do not feel the need to control their data and show no interest in data rights~\cite{reis2018patients, alizadeh2020gdpr, kaushik2021know, jilka2021terms, duckert2022protecting, sahqani2021co, nunes2020futures}. In fact, a survey shows that, on average, users have a neutral willingness to exercise data rights~\cite{sorum2022gender}. This is counterintuitive since people generally value their privacy~\cite{kokolakis2017privacy, acquisti2015privacy}. We explore the reasons behind this lack of desire for data rights:
\newline 
\indent\indent (i) \textbf{Trust:} Trust in companies holding personal data impacts users' perceptions of data rights~\cite{jilka2021terms, sahqani2021co}. Users who have high trust in a company are less likely to feel the need to read privacy policies or to be informed about the company's data practices~\cite{jilka2021terms, sahqani2021co}.
\newline 
\indent\indent (ii) \textbf{Indifference:} 
Some users exhibit indifference or insensitivity towards the use of their data, which diminishes their interest in data rights~\cite{reis2018patients, presthus2019consumer, kaushik2021know, eiband2018bringing}.
As an example, a small fraction of patients in a study (about 5\%) did not find it problematic for others to access their health information at any time, thus showing no interest in their data rights~\cite{reis2018patients}.
Other individuals perceive that personalized ads and direct marketing do not harm their privacy significantly; hence, they do not aspire to exercise their data rights~\cite{presthus2019consumer}: ``\textit{I will most likely not pursue this [the right to erasure]. From the advertisements that Facebook continuously shows me, I draw the conclusion that they do not really know too much about me as a person}.'' 
Further, a minority of participants in a user experiment (2 out of 16) were not concerned about the transparency of a systems~\cite{eiband2018bringing}.
\newline 
\indent\indent(iii) \textbf{Privacy fatigue:} Another key reason for users' lack of desire for data rights is the perceived burden and cost of exercising these rights~\cite{alizadeh2020gdpr, duckert2022protecting, sailaja2019living, farke2021privacy, vitale2019keeping}. Some users deem the explanations of automated decision-making boring and irrelevant~\cite{kaushik2021know}. The constant need to exercise data rights is particularly undesirable, with very few expressing a wish for such continuous engagement~\cite{alizadeh2020gdpr, zhang2021facial}. Even though some individuals recognize the importance of data rights, they find that active engagement leads to privacy fatigue and annoyance~\cite{duckert2022protecting, sailaja2019living, vitale2019keeping}. This burden leads some users to prefer delegating their data rights to others, such as automatic tools or third parties, rather than managing their data directly~\cite{vitale2019keeping, marky2020all}.
Interestingly, even with accessible online tools that simplify the exercise of data rights, some users still perceive the process as too time-consuming~\cite{farke2021privacy, vitale2020data}: ``I have better things to do with my time, frankly, than to be reviewing this.''
\newline 
\indent\indent (iv) \textbf{Fear and low understanding:} 
Additionally, some users may not desire data rights due to heightened fear from learning more about personal data processing, or due to a lack of understanding about the benefits of data rights~\cite{alizadeh2020gdpr, presthus2021three}. For instance, the right to data portability is often undervalued due to a lack of understanding~\cite{presthus2019consumer, presthus2021three, sailaja2021human}.
\newline
In conclusion, while there is significant interest in data rights, there is also substantial evidence of user apathy towards them.
\smallskip

\noindent \textbf{(U3) Do users exercise data rights in practice?}
\smallskip

\indent (U3.1) \textit{Attitude vs. actual behavior}
Even if users have a significant desire for data rights, they rarely exercise them in practice~\cite{presthus2019consumer, kulyk2020has, presthus2021three, urban2019your, meng2021owning}. 
For instance, a survey indicates that while many express the desire to understand how companies utilize their cookies, few actually read the privacy policy to obtain this information~\cite{kulyk2020has}. 
A striking disparity was observed with GDPR implementation; many users who previously expressed strong intentions to exercise data rights did not follow through~\cite{presthus2021three}. 
Specifically, for the right to delete, about 40\% of the users claimed that they will ``definitely'' use this right before enforcing the GDPR, but the following survey conducted after 1 and 2 years shows that only about 15\% had exercised the right~\cite{presthus2021three}. 
This incongruity is consistent with the well-known ``privacy paradox''---a discrepancy between privacy concerns and actual behavior~\cite{kokolakis2017privacy, acquisti2015privacy}. 
\newline 

\indent (U3.2) \textit{Users' non-exercise of data rights}.
The reality is that many users do not exercise GDPR data rights. It is common for people to quickly scroll through without actually reading privacy policies~\cite{liao2020measuring, presthus2019consumer, meng2021owning, barrett2018critical, beierle2021tydr, tiktok, obar2020biggest, phillips2021uk}. User experiments on a tracking app and a social networking service found that over 90\% of participants spent less than 30 seconds on the privacy policy page~\cite{beierle2021tydr, obar2020biggest}, and it seems that only 1\% of people read the entire text of privacy policies~\cite{presthus2019consumer}.
\newline 
Moreover, the lack of exercising rights is not limited to the right to information; it extends to other rights, including the right to erasure. Most Europeans (70\% of survey respondents~\cite{mangini2020empirical}) have never applied for data deletion, and only 1.13\% frequently delete their cookie data~\cite{papadopoulos2019cookie}. 
Service providers also report a lack of requests for data access and erasure~\cite{alizadeh2020gdpr, urban2019your, phillips2021uk}. Some companies claim there had been no user access requests and less than ten profile deletion requests over a full year, contrary to their expectations~\cite{urban2019your, phillips2021uk}. 
Additionally, data portability is infrequently utilized; only 7\% of users surveyed have exercised this right~\cite{presthus2019consumer}. Therefore, it is not common for users to exercise GDPR data rights. 
\newline 

\indent (U3.3) \textit{Users' exercise of data rights}.
However, there are instances where users actively utilize specific GDPR data rights, particularly the right to object. About 66.3\% of U.S. consumers have opted out of receiving emails from e-commerce retailers, thereby avoiding direct marketing~\cite{willis2021trust}.
\newline 
While rights to information and erasure are not commonly exercised, users become more proactive when the data in question is highly valued. For example, where data is genetic information, some users actively exercise the rights to information and access~\cite{grandhi2022pass}.
The users closely monitor updates to privacy policies and terms of service, download their genetic data for safekeeping, and are prepared to delete their information if issues arise.
\newline 
In contexts involving their children (e.g., smart homes), users can also pay more attention to privacy policies in smart devices surveilling their homes~\cite{sun2021child}; certain parents consider privacy policies crucial to their children's safety when selecting smart devices~\cite{sun2021child}.
Furthermore, some users regularly examine and delete personal data collected by smart devices related to their daily lives~\cite{lau2018alexa, meng2021owning, chalhoub2021did}. 
\newline 
It's important to note that, as highlighted in U3.2, rights to information and erasure are generally not actively exercised. However, these instances suggest that user behavior towards data rights varies depending on the type of data and context. Thus, we should avoid blanket statements about data rights exercise, as user responses can be heavily influenced by individual circumstances and perceived value of the data involved.
\smallskip

\noindent \textbf{(U4) Do data rights benefit users?}
\smallskip

\indent (U4.1) \textit{Effect on user perception}.
Privacy concerns and uncertainties that users have when using a service or product can degrade their service/product experience. Therefore, if data rights can reduce perceived privacy concerns and uncertainties, it may encourage users to choose companies that support their data rights, which ultimately benefits users. On the contrary, if data rights have a negative impact on user perception, users may prefer non-compliant companies, contradicting the original intention of privacy laws. 
\newline
Data rights have a significant influence on users' perceived privacy concerns and uncertainties. Specifically, providing users with higher, simpler control over their data can mitigate their perceived privacy concerns, risks, and uncertainties~\cite{willis2021trust, herder2020privacy, farke2021privacy,kulyk2020has}. A user experiment, for instance, demonstrates that showing users the personal information they have shared can decrease perceived privacy risks compared to merely viewing a privacy policy~\cite{herder2020privacy}. When the system further allowed users to delete and modify their data, perceived privacy concerns and risks decreased. 
\newline 
However, the same experiment~\cite{herder2020privacy} questions the assumption that granting complete access rights always benefits users. When the system permitted users to access inferred data, which they did not provide but was inferred by the company from the provided personal information, users felt an increased sense of privacy risk.
\newline 
Moreover, the implementation method of data rights significantly impacts users' perceived privacy concerns. For instance, one study reveals that merely justifying data collection to users does not significantly affect perceived uncertainty~\cite{liu2022protecting}. Yet, the same study reports that perceived risk decreases when users encounter a privacy statement including phrases such as ``strictly protect your privacy.'' Another user survey suggests that the more effective a user considers a given privacy policy, the less privacy risk they perceive~\cite{paul2020privacy}. These findings suggest that the relationship between the right to information and perceived privacy risk heavily relies on how a privacy policy is presented, including word choice. 
\newline 
Moreover, there is evidence indicating that the right to avoid automated decision-making does not significantly improve user perception regarding fairness, accountability, trustworthiness, and feelings of control~\cite{vaccaro2020end}.

\indent (U4.2) \textit{Effect on users' privacy-seeking behavior}.
Secondly, a study indicates that reminding users of GDPR data rights can curb data sharing behavior~\cite{palinski2021pay}. The research examined how many users would install an app and permit a system access request after being informed of data rights at the outset of the experiment. The result suggests fewer users shared their personal information with the application system when reminded of GDPR data rights.
\newline 
However, this observation contradicts previous findings~\cite{brand2013misplaced,mothersbaugh2012disclosure, chanchary2015user} suggesting that providing users with data control can paradoxically lead to more data disclosure, a phenomenon known as the Peltzman effect where people tend to take riskier actions under security measures. While the exact reason for this contradiction is not clear, one hypothesis is that the mention of GDPR itself stimulated privacy-seeking behavior among users in the study~\cite{palinski2021pay}. The only major difference across these studies is the mention of GDPR, and it's possible that the priming effect of the term GDPR outweighed the Peltzman effect. Relatedly, one study~\cite{chanchary2015user} reports that granting data control to users did not significantly alter their data-sharing behavior; most participants maintained their willingness not to share personal information even after being given data control. Furthermore, evidence shows that users are more inclined to disclose their health data when given a reason for data collection that highlights the benefits of the product~\cite{becker2020s}.
In conclusion, further research is necessary to understand the impact of GDPR data rights on data disclosure behavior.






%% file: 005_Company.tex
\section{Companies' perspective}
\label{sec:company}

We now turn our attention to eight key questions concerning companies' understanding, attitudes, implementation, challenges, and benefits pertaining to users' data rights.

\smallskip
\noindent \textbf{(C1) Are companies informed about and understand data rights?}

A significant number of companies demonstrate a limited understanding of data rights~\cite{zanker2021gdpr,nguyen2021share,norval2021data}. A user survey of businesses across eight EU countries revealed that only approximately 50-65\% of participants correctly answered questions about the right to access, the right to erasure, and the right to object~\cite{zanker2021gdpr}. Additionally, some app developers mistakenly believe that the GDPR is inapplicable to them because their primary market is outside the EU~\cite{nguyen2021share}. Similarly, several startups misinterpret the GDPR, asserting that certain data rights cannot apply to their key technologies~\cite{norval2021data}. Contrary to their beliefs, the GDPR applies to all businesses collecting personal data from citizens of European Economic Area (EEA) countries, irrespective of their primary market and business sectors. Notably, studies by Zanker et al. \cite{zanker2021gdpr} and Nguyen et al. \cite{nguyen2021share} conducted in early 2020 and early 2021, respectively, highlight the persistent deficiency in companies' understanding of GDPR data rights even 2-3 years after its implementation.

\smallskip
\noindent \textbf{(C2) What is companies' attitude toward complying with data rights?}

\indent (C2.1) \textit{Indifference to data rights}.
Given the prevalent lack of awareness of GDPR data rights, one might expect indifference towards users' data rights. Indeed, there is substantial evidence for such indifference. For instance, despite the rising interest in applying the right to information to machine learning (ML) systems~\cite{goodman2017european}, no ML practitioners express concerns about this issue in an interview~\cite{hopkins2021machine}. 
ML algorithms are often deemed `black boxes,' complicating the enforcement of the right to information. Furthermore, even though the right to erasure implementation within an ML model is widely acknowledged as challenging, many practitioners are unaware of this issue and have not considered the implications of the right to erasure in their practices of applying ML~\cite{hopkins2021machine}. Likewise, an experiment reveals that of 448 developers, 334 ignored an email alerting them to incorrect implementation of data rights in their apps~\cite{nguyen2021share}. Surprisingly, some small-scale business stakeholders openly express that GDPR compliance is unnecessary, assuming their businesses are too small to attract regulatory attention~\cite{company_survey} ($<$ 500 employees). 

Further, in an interview, some service providers loaded the responsibility of implementing data rights to others: ``\textit{No, not me! It's the designers maybe...}''~\cite{alkhatib2020privacy}. 
There is also skepticism about the right to data portability from the perspective of service providers, e.g., some developers believe that the right to data portability brings them no benefit and do not expect that they will receive data portability requests because there are few direct competitors~\cite{norval2021data}.
Overall, these testimonies underscore that many companies are still indifferent to implementing data rights and have skepticism towards certain data rights.

\indent (C2.2) \textit{Willingness to implement data rights.}
Not all developers disregard users' data rights. Many service providers emphasize the importance of enforcing them~\cite{huth2020empirical, zieni2021transparency, alkhatib2020privacy, company_survey, lyons2021conceptualising, luusua2021nordic, grundstrom2018transforming}. They highlight the growing user awareness of their rights and believe the exercise of data rights will gain prominence with the adoption of GDPR~\cite{huth2020empirical,zieni2021transparency}.
Most European business leaders (about 86\%) recognize the importance of GDPR compliance~\cite{company_survey}, a figure significantly higher than average individuals fully understanding GDPR data rights (50-65\%)~\cite{zanker2021gdpr}. This indicates that service providers can appreciate the importance of GDPR, even with a partial understanding of the law. Reasons for compliance include fear of fines and respect for users' data rights~\cite{company_survey}.
Likewise, practitioners knowledgeable in ML showed appreciation for the right to get a human review in automated decision-making and stated the importance of ML's transparency~\cite{lyons2021conceptualising, luusua2021nordic}.
\newline 
Evidence of service providers' commitment to support users' data rights can also be seen in practice. Online developer forums frequently discuss how to design and compose privacy policies or consents for their apps~\cite{li2021developers}.
As a result, we can also see that many service providers feel that implementing user data rights in their systems is essential. 
Consequently, there is an opposing force to C2.1, as a significant number of service providers also recognize the importance of implementing user data rights within their systems.

\smallskip
\noindent \textbf{(C3) Have companies made internal changes to support data rights?}

Adherence to the GDPR data rights necessitates significant effort from companies, prompting various internal measures such as employee training, consultation procurement, and data management process modifications~\cite{zanker2021gdpr,company_survey,hopkins2021machine,huth2020empirical}. A survey reveals that approximately 91\% of European companies have conducted at least one employee training session on this subject~\cite{zanker2021gdpr}. Numerous companies report engaging in GDPR compliance consultation services~\cite{company_survey}. In preparation for the GDPR, most companies have incurred costs between \texteuro1,000 and \texteuro50,000, predominantly for employee education and consultancy services~\cite{zanker2021gdpr, company_survey}.
\newline 
However, the effectiveness of the existing training is in question, given the incomplete understanding of GDPR data rights among many developers despite widespread training (see C1). This lack of understanding may stem from the inadequate number of training sessions provided by companies. A survey suggests that most companies only held one training session on the GDPR~\cite{zanker2021gdpr}, indicating that a greater number of sessions may be required to adequately educate employees about data rights.
\newline 
Several companies have implemented substantial changes to their data management processes to comply with data rights~\cite{hopkins2021machine, huth2020empirical,jantti2020studying}. An interview with a publicly listed ML company revealed significant changes to their data handling practices over six months, guided by their legal team to ensure GDPR compliance~\cite{hopkins2021machine}. Enterprise Architecture Management (EAM), which provides companies with strategies for efficient development, has played a crucial role in implementing data rights~\cite{huth2020empirical}. For instance, some EAMs have defined procedures for developers to follow when implementing data rights and have specified the correct communication channels for notification.
\newline 
Many companies maintain logs of all requests for user data rights, for example, of their deletion records~\cite{bertram2019five}. A company developed a research tool to understand how its users consider transparency~\cite{grundstrom2018transforming}.
Thus, there is substantial evidence that many companies are striving to implement GDPR data rights internally.
\newline 
Nevertheless, some companies have opted not to implement data rights. One health tech startup, for example, took legal advice against implementing the right to erasure as they had no plans to allow patients to exercise this right~\cite{norval2021data}.
\newline 

\noindent \textbf{(C4) Do services operated by companies enable users to exercise data rights?}

\indent (C4.1) \textit{Right to information}.
A plethora of studies have scrutinized the compliance of privacy policies with the GDPR across various platforms such as web, mobile applications, and IoT~\cite{akanfe2020assessing, hatamian2021privacy, liu2021have, zaeem2020effect, sanchez2021automatic, vanezi2021complicy, oh2021will, strzelecki2020consumers, brown2020whose, fan2020empirical, milkaite2020child, urban2019your, vallina2019tales, mohan2019analyzing, paul2018assessing, sun2020quality, sanchez2019cookie, amos2021privacy, liao2020measuring, zhang2020does, sharma2021privacy, sundareswara2021large,tesfay2018privacyguide, o2021implementing, benjumea2020assessment, akanfe2020design, davis2019contracting, benjumea2019privacy, boniface2019security, hu2019characterising, mehrnezhad2021caring, khalil2018unbearable}.
These studies employ manual analyses or automated tools, such as Machine Learning (ML) or Natural Language Processing (NLP), to scrutinize privacy policies. Although methodologies and evaluation criteria vary across studies, the consensus is that few privacy policies currently meet all GDPR requirements.
\newline 
A noticeable disparity exists in the compliance rates for individual GDPR requirements. Over 90\% of policies outline the collection of personal data and its purposes, though some instances of incomplete information have been reported~\cite{liu2021have, zaeem2020effect, fan2020empirical, zhang2020does, khalil2018unbearable}. However, the majority of policies do not describe data rights like access and erasure~\cite{liu2021have, boniface2019security} even though European websites tend to state more extensive data rights than websites from other countries~\cite{woocomparing}. Furthermore, many policies struggle to accurately inform users about the flow and transfer of data~\cite{vallina2019tales, zhang2020does, subahi2018ensuring, kampanos2021accept, mehrnezhad2021caring}. A significant difference in compliance rates across websites in various countries is also found; European websites tend to have a higher compliance rate than others~\cite{woocomparing, hu2019characterising}.
\newline 
Surprisingly, some services do not provide a privacy policy at all~\cite{liao2020measuring, sharma2021privacy, vallina2019tales, oh2021consent, o2021implementing, mehrnezhad2021caring, mehrnezhad2020cross}. The availability of privacy policies varies significantly across platforms. For instance, most websites (4,566 out of 5,045 surveyed) provide a privacy policy~\cite{oh2021consent}. Yet, a mere 16\% of 6,843 analyzed pornographic websites have a privacy policy~\cite{vallina2019tales}. Similarly, only 62\% of Google Assistant actions, which serve as voice-assistance applications, provide a privacy policy~\cite{liao2020measuring}. Moreover, more than half of Greek and UK and more than one-third of German websites did not include a cookie notice~\cite{kampanos2021accept, krisam2021dark}, and 12 out of 30 fertility apps analyzed did not prompt privacy notices~\cite{mehrnezhad2021caring}. 
\newline 

\indent (C4.2) \textit{Rights to access, data portability, erasure, and object}.
A multitude of studies have delved into the extent to which systems uphold the right to access and data portability, typically by issuing a direct request~\cite{turner2021exercisability, veys2021codesign, tolsdorf2021case, bufalieri2020gdpr, kroger2020app, sorum2020dude, urban2019your, urban2019study, wong2018portable, zwiebelmann2021data}.
These studies generally reveal that the majority of requests are honored within the one-month GDPR deadline.
Similarly, an interview indicated that over 90\% of service providers processed all requests for access and erasure within a month~\cite{phillips2021uk}.
However, compliance rates can be low in specific sectors~\cite{kroger2020app, urban2019study}; for instance, a 2019 study showed that only 41\% of popular German mobile apps provided a copy of personal data within a month~\cite{urban2019study}.
\newline 
Moreover, past studies frequently observe that the data procured through the access right are incomplete~\cite{veys2021codesign, tolsdorf2021case, urban2019study, kroger2020app, pins2021alexa}; the data may not include certain personal information that users supplied.
Furthermore, while many systems allow users to download their personal data in a machine-readable format, few provide the data import features required for data portability~\cite{kuebler2021right, turner2021exercisability}. There are examples of systems rejecting users' requests to download their data without providing a justification~\cite{zwiebelmann2021data}.
\newline 
Evidence suggests that the right to erasure is often unsupported, with many instances where companies refused users' requests~\cite{presthus2021analysis}. A study further reveals that companies disregarded users' opt-out requests for direct marketing~\cite{da2018comparison}. Consequently, many systems fail to fully uphold data rights.
\newline
Interestingly, many practitioners might dispute this fact. Indeed, a survey revealed that over 80\% of practitioners believe their organizations are mostly or fully GDPR-compliant~\cite{company_survey}. Likewise, in an interview, some developers confidently claimed their companies possess adequate resources to meet GDPR requirements~\cite{jantti2020studying}. These assertions, juxtaposed with evidence of companies' actual compliance status, suggest a potential lack of awareness about their GDPR compliance. In fact, some respondents expressing full compliance lacked confidence when asked specific compliance questions~\cite{company_survey}.
\newline

\noindent \textbf{(C5) Is it technically and managerially feasible for companies to implement data rights?}

Numerous technical and managerial challenges can make it virtually impossible for businesses to implement data rights. For instance, some suggest that, for efficiency, companies should deploy a system managing all applications processing personal information, as users can make daily requests to control their data~\cite{huth2020empirical}. However, this would be challenging for applications run by independent departments~\cite{huth2020empirical}. Furthermore, budget constraints can impede the implementation of data rights, as companies need to invest significantly in employee training and IT solutions for data management~\cite{chal-ben, sirur2018we}.
\newline 

\indent (C5.1) \textit{Right to erasure}.
Among the eight GDPR rights, the right to erasure was most frequently cited as challenging by developers~\cite{norval2021data,aydin2020gdpr,mangini2020empirical,boenisch2021never,hopkins2021machine,huth2020empirical,ciordas2020strategies, poritskiy2019benefits}.  
Removing user data, especially from backups and archives, appears to be a complex, costly, and unclear process~\cite{huth2020empirical,ciordas2020strategies, mangini2020empirical}.
When backups need to be restored, previously deleted data may resurface, suggesting a failure in the erasure process.
Moreover, larger enterprises tend to struggle more with this right due to the sheer volume of data~\cite{poritskiy2019benefits}.
Some even admit that applying this right in their systems is impossible~\cite{aydin2020gdpr}. 

Confusion amplifies when considering ML~\cite{ boenisch2021never, boenisch2021never, hopkins2021machine}.
There is a lack of clear guidelines on whether to remove deleted data from all model training, test, and validation sets, and whether to delete the model itself upon receiving an erasure request from users.
Another aspect is a conflict between data rights and technologies' goals. 
Representatively, blockchain companies suffer from this problem~\cite{norval2021data}.
Given the immutability of blockchain technology, erasing data stored in the system is inherently impossible. Many blockchain companies discuss this issue in their privacy policies, with some stating outright that exercising this right is impossible~\cite{sauglam2020data}. However, a few claims they can delete personal information from their blockchain systems, but details on how this is achieved are typically undisclosed~\cite{sauglam2020data}.
\newline 
Nonetheless, not all companies agree with these challenges; some consider implementing the right to erasure to be a trivial matter~\cite{norval2021data}.
\newline 

\indent (C5.2) \textit{Right to information and access}.
Companies often find the rights to information and access challenging to implement~\cite{nguyen2021share, aydin2020gdpr, urban2019your}.
Firms must provide accurate information about data processing to users, necessitating a comprehensive understanding of data flows and processing mechanisms.
However, many companies struggle with this, especially when third parties involved in online advertising also process the data. Full awareness of how these third parties handle user data can be a formidable task. Several studies reveal that most developers grapple with this lack of knowledge~\cite{nguyen2021share, urban2019your, li2018coconut}.
\newline 
Moreover, the right to information poses a challenge in ML systems~\cite{hopkins2021machine, goodman2017european, dhanorkar2021needs, bhatt2020explainable}.
The right to information necessitates that corporations disclose the underlying logic of algorithms that may have consequential or legal effects on users. This disclosure, however, may not be feasible due to the inherent ``black box'' nature of ML algorithms. There are extant techniques, such as LIME~\cite{ribeiro2016should}, devised to demystify this issue, yet the resultant explanations are often technically complex and inaccessible to users lacking comprehensive ML knowledge~\cite{dhanorkar2021needs, bhatt2020explainable}. This complexity often leads engineers to use these interpretability methods primarily to enhance ML system efficiency, deviating from the original intent of explicating algorithms to users~\cite{bhatt2020explainable}.
\newline 
Moreover, tensions may emerge between data rights and corporate interests, as ML algorithms can be considered proprietary to a company. This poses a critical question: To what extent should companies be obligated to reveal their algorithms~\cite{dhanorkar2021needs}? A high degree of ML transparency could paradoxically infringe on user privacy by potentially revealing user identities embedded within the training dataset\cite{dhanorkar2021needs,bhatt2020explainable}.
\newline 
Additionally, developers identified user identification as a challenge in enforcing the right to access in an interview~\cite{urban2019your}.
This identification is crucial to ensure that only data owners can access their personal information.
However, strong user authentication might require them to provide sensitive data, such as an ID card, which could infringe on privacy, contradicting the original intent of the rights.
Companies should carefully consider the level of identification before granting users the right to access~\cite{urban2019your} and should ensure robust security in the identification process.
However, currently, this is often not the case, a point which will be elaborated on in C7.
\newline 

\indent (C5.3) \textit{Right to data portability}.
Furthermore, implementing the right to data portability, which should enable data transfer between companies, presents a challenge due to data interoperability issues arising from heterogeneous data structures among companies~\cite{norval2021data}.
\newline

\indent (C5.4) \textit{Right to avoid automated decision-making}.
The requirement for human review in decision-making poses several challenges, including ``decision fatigue, complacency, bias, and scalability''~\cite{lyons2021conceptualising}.
\newline 

\indent (C5.5) \textit{Disparity}.
The preceding discussion implies a broader implication: Data rights may disproportionately impact certain companies.
The degree of difficulty and challenges in implementing data rights varies across companies.
For instance, large companies may struggle with the sheer volume of data, while budget constraints and the need for robust management systems might burden smaller or newer companies more heavily.
These diverging difficulties suggest an unequal burden across the business spectrum. We refrain from determining whether large or small companies bear a greater burden, as it is not a straightforward matter due to these conflicting challenges.
Consequently, current privacy laws could inadvertently disadvantage certain companies, thus impeding fair competition. This overlap between privacy and competition laws may need careful consideration.
\newline 

\noindent \textbf{(C6) Does the user process of exercising data rights have high usability?}

\indent (C6.1) \textit{Usability challenges}.
The process for users to exercise data rights is generally considered to have poor usability.
Consider privacy policies, for instance. Numerous studies have concluded that these policies are typically too verbose and complex for the average person, as evidenced by various readability metrics, including word count and Flesch scores~\cite{liao2020measuring, jilka2021terms, urban2019your, anikeev2021privacy, oh2021consent,milkaite2020child, javed2020south, wett2020decision, renaud2018privacy,sanchez2019cookie, habib2019empirical, bascur2021ethical, valtysson2021co, khalil2018unbearable}.
On average, these policies contain about 2000 words and are written at a 13th-grade reading level, presenting substantial comprehension challenges for many users.
Consistent with this, users frequently express frustration with the difficulty of understanding privacy policies~\cite{liao2020measuring, presthus2019consumer, barrett2018critical,liu2021have}. The implementation requirements according to the GDPR data rights can degrade the usability of platforms~\cite{wagner2020regulating, valtysson2021co}.
\newline 
These usability issues, along with dark patterns that result in discouraging users' exercise of data rights, extend beyond the right to information and impact other rights such as the rights to access, erasure, data portability, and objection~\cite{prior2020parents,kuntsman2019digital,ostheimer2019privacy,gunawan2021dark,mathur2019dark,habib2019empirical,habib2020scavenger, veys2021codesign, pins2021alexa, vanezi2019gdpr, gutmann2019fight,zwiebelmann2021data, darkpatters}. A platform allows users to access their data only when they email them in a specific format~\cite{vanezi2019gdpr}. 
Numerous studies also point to the difficulty of understanding downloaded data due to its volume, disorganization, use of jargon, and a lack of description~\cite{veys2021codesign, 138, pins2021alexa, zwiebelmann2021data}.
Additionally, research reveals that users often struggle to find ways to delete their personal data from apps and services~\cite{prior2020parents,kuntsman2019digital,ostheimer2019privacy}.
For example, disproportionate amounts of time to find a ``delete"~\cite{ostheimer2019privacy} or ``opt-out''~\cite{habib2020scavenger} button or the need to call the service team to do so~\cite{kuntsman2019digital}.
Indeed, the absence of straightforward account deletion options is a pervasive issue in online services, affecting about 77\% of them~\cite{gunawan2021dark}.
Overall, the evidence suggests that exercising data rights is a nontrivial task for many users.
\newline 

\indent (C6.2) \textit{Usability Successes}.
Despite the above challenges, there are instances where users have found the process of exercising their rights straightforward, particularly in the cases of the rights to access and erasure~\cite{alizadeh2020gdpr, mangini2020empirical,veys2021codesign, habib2020scavenger}.
For example, a survey shows that 58\% of users who had previously requested companies to erase their personal information found the process straightforward~\cite{mangini2020empirical}.
Furthermore, a study found that users had no difficulty in requesting a copy of their personal information from a loyalty card provider~\cite{alizadeh2020gdpr}.
Certain websites, like Amazon, Facebook, and Google, provide online portal tools to facilitate the rights request process, eliminating the need for users to contact the companies directly~\cite{veys2021codesign}.
Consequently, the usability of exercising data rights varies significantly across different applications and data rights types.
\newline 

\indent (C6.3) \textit{High-usability Data Rights Tools}. 
Given the usability issues associated with exercising data rights, the development of user-friendly data rights tools has been a natural response.
Developers and researchers proposed and launched many tools and services that allow users to easily make use of data rights~\cite{sahqani2021co, liu2021have, kumar2020haystack, fernandez2021effects, seymour2020inform, railean2018lite, bbc, myactivity, karegar2020dilemma, karegar2018helping, bruggemeier2022perceptions, pins2021alexa, pilton2021evaluating}. 
While several studies validate that such tools can increase the effectiveness of data rights~\cite{sahqani2021co, liu2021have, farke2021privacy, kumar2020haystack, fernandez2021effects, seymour2020inform, railean2018lite, bruggemeier2022perceptions}, we can also find some testimonies that raise doubts about their effects~\cite{urban2019your, farke2021privacy, sailaja2021human, karegar2020dilemma, alizadeh2020gdpr, bowyer2018understanding, pins2021alexa}. Given this, the effects of ``high usability'' tools are still unclear.
\newline 

\indent (C6.3.1) \textit{Effectiveness.}
Empirical research suggests that high-usability data rights tools can effectively assist users in asserting their data rights.
Emphasis is on tools that aim to mitigate the common challenge of understanding privacy policies~\cite{sahqani2021co, liu2021have, kumar2020haystack, fernandez2021effects, seymour2020inform, railean2018lite, bbc, myactivity, karegar2020dilemma, bruggemeier2022perceptions, pilton2021evaluating}.
They commonly employ visualization, summarization, and user engagement strategies to simplify data processing comprehension.
\newline 
For instance, designs using visualization techniques display icons representing collected data or diagrams illustrating data flows, aiding intuitive understanding~\cite{seymour2020inform,railean2018lite,fernandez2021effects}.
In contrast, some tools prioritize brevity and highlight crucial points in privacy policies, presenting users with a summarized version~\cite{liu2021have,kumar2020haystack, yao2019defending, pilton2021evaluating}.
Users concur that these techniques--visualization and summarization--facilitate understanding of policy content compared to conventional privacy policies.
Moreover, research suggests the significant role of color in enhancing comprehension, with vibrant hues being particularly effective.
Such approaches could also be applied to improve the readability of data obtained via the right to access~\cite{veys2021codesign, pins2021alexa}.
\newline 
Another category of tools promotes user engagement as a strategy to increase awareness of policy content~\cite{karegar2020dilemma, karegar2018helping}.
This is predicated on the established theory that active learning fosters more effective comprehension~\cite{bornstein2014interaction, piaget2013mechanisms, vygotsky1991genesis}.
Empirical evidence confirms enhanced user awareness when users engage actively with policy content via ``Drag-and-Drop'' and ``Question-and-Answer'' features.
User experiments also reveal a higher likelihood or willingness of data rights assertion when interacting with a chatbot~\cite{bruggemeier2022perceptions, musto2020towards}.
In conclusion, ample evidence supports the efficacy of designs employing visualization, summarization, or user engagement in improving the effectiveness of data rights.
\newline 

\indent (C6.3.2) \textit{Ineffectiveness.}
Despite promising results, some studies question the long-term effectiveness of these high-usability designs.
Research~\cite{karegar2020dilemma} suggests that although user-engagement designs are initially effective, this impact diminishes with repeated exposure.
Moreover, while visualization and user engagement can increase user awareness, they may inadvertently reduce willingness for data rights due to visual fatigue or annoyance caused by repeated exposure.
This is particularly relevant considering privacy fatigue is a primary factor in data rights apathy (see U2.2).
Furthermore, visualization does not always prove effective in elucidating complex concepts. In a specific industrial instance, employing visualization to interpret an inherently intricate and technical ML algorithm paradoxically engendered more confusion than its textual counterpart~\cite{dhanorkar2021needs}.
\newline 
Additionally, a user study found that despite 75\% of participants finding a new data rights tool helpful, only 37\% expressed willingness to use it to exercise their rights~\cite{farke2021privacy}. Further evidence shows that most users, while satisfied with a proposed data rights tool, do not perceive the regular need to exercise their rights using the tool~\cite{pins2021alexa}.
This discrepancy becomes more pronounced when considering that expressed willingness does not always translate into action, as discussed in U3.1.
Furthermore, users struggled to use an online tool designed to exercise the right to data portability, primarily due to a lack of understanding of the data right itself~\cite{sailaja2021human}.
\newline 
The ineffectiveness of such tools is further observed in real-world scenarios.
For instance, an interview with developers revealed that companies providing online portals for consumers to exercise data rights did not receive more rights requests than those offering only offline tools~\cite{urban2019your}.
As a result, many companies do not prioritize the development of online tools. 
\newline 
Interestingly, many users do not expect companies to provide easy-to-use online interfaces to exercise data rights.
In an interview, while some users, particularly those more interested in their data rights, express a desire for online data rights tools, the majority view traditional means such as email or phone as the most appropriate channels to claim their access rights~\cite{alizadeh2020gdpr}.
Some users even express a preference for offline data rights exercises, favoring direct interaction with a responsible party~\cite{bowyer2018understanding, nurgalieva2019information}.  
Furthermore, an evaluation of a purportedly ``high-usability'' data dashboard~\cite{vitale2020data} revealed unfavorable responses. A subset of participants demonstrated no inclination or necessity to utilize it, some voiced objections to the very notion of ``high-usability'' dashboards, and others dismissed data management as inconsequential.
\newline 
In summary, despite their high usability, the real-world effectiveness of these tools remains uncertain.
This is not to say these tools are ineffectual, but rather that the evidence indicates that high usability alone does not necessarily enhance the current effectiveness of data rights.
A multi-faceted approach addressing all dimensions influencing data rights effectiveness is thus needed in our society.
\newline 

\noindent \textbf{(C7) Are data rights being implemented without creating new technical flaws or threats?}
\newline 
The implementation of data rights should not expose users to additional threats.
Current practices, however, often fail to satisfy this criterion.
Specifically, research indicates a risk of leaking personal information to unauthorized third parties when users exercise their right to access~\cite{bufalieri2020gdpr, cagnazzo2019gdpirated, di2019personal, boniface2019security}.
Companies must verify the identities of those making data rights requests to ensure they are the actual data owners.
A flawed identification process can result in the leakage of sensitive personal information to unauthorized individuals.
Several studies reveal system vulnerabilities susceptible to impersonation via counterfeit emails~\cite{bufalieri2020gdpr, cagnazzo2019gdpirated, di2019personal}.
These studies demonstrate how easily an attacker can deceive service providers by using an email address similar to the original one, with success rates ranging from approximately 30\% to 70\%~\cite{bufalieri2020gdpr, cagnazzo2019gdpirated, di2019personal}.
Some systems do not even implement email encryption, despite requiring users to submit identification documents via email~\cite{bufalieri2020gdpr, boniface2019security}.
Hence, severe threats often arise when users exercise their right to access.
\newline

\noindent \textbf{(C8) Do data rights produce positive effects for complying companies?}

\indent (C8.1) \textit{Efficient data use.}
Several studies contend that the implementation of GDPR data rights significantly enhances companies' data processing awareness, leading to improved user data management and increased competency in utilizing  data~\cite{hopkins2021machine, mangini2020empirical, chal-ben, poritskiy2019benefits, bennett2018european, sirur2018we}.
The growing responsibility of companies in the information economy can also foster better decision-making and risk assessment~\cite{chal-ben}.
Large companies, in particular, seem to strongly recognize the benefits of improved data management processes~\cite{poritskiy2019benefits}.
\newline 
However, to assess whether data rights lead to efficient data use, it is important to consider not only companies' data processing but also users' data sharing behaviors.
If data rights cause users to be more reticent about sharing their data, it could compromise the efficiency of data use by companies, despite improvements in internal data management processes.
As discussed in U4.2, the evidence on this matter is conflicting~\cite{brand2013misplaced,mothersbaugh2012disclosure, chanchary2015user,palinski2021pay,becker2020s}.
While some studies suggest that empowering users may generally incentivize them to disclose more personal data~\cite{brand2013misplaced,mothersbaugh2012disclosure, chanchary2015user,becker2020s}, one report indicates that explicitly linking user data rights with the term GDPR can have the opposite effect~\cite{palinski2021pay}.
Therefore, in contrast to previous studies, we posit that it remains uncertain whether GDPR data rights enhance data use efficiency within companies. This area warrants further research.
\newline 

\indent (C8.2) \textit{Positive corporate image.}
There is evidence of an increased trust of users in companies that foster data rights \cite{mangini2020empirical,willis2021trust, mohallick2018privacy, herder2020privacy}.
Survey results suggest that service providers can bolster their credibility by offering options to modify/delete personal information and maintaining transparency about data processing~\cite{mangini2020empirical,mohallick2018privacy,herder2020privacy}.
A separate survey indicates that U.S. consumers place greater trust in companies that voluntarily extend GDPR data rights to non-European users~\cite{willis2021trust}.
These findings suggest that user empowerment can enhance a company's image.
\newline 
However, the relationship between data rights and trust is multifaceted.
The extent to which data rights can rehabilitate a company's image, if it is initially negative, remains uncertain.
Under conditions of low trust, users may question the data rights themselves~\cite{presthus2021three,barrett2018critical,da2018information, chanchary2015user, pew_cont}, making it challenging for companies to anticipate the positive effects of endorsing data rights.
For instance, some interviewees express skepticism about whether companies would actually delete their personal information upon request~\cite{da2018information}.
Another study reveals a perception that companies deliberately obfuscate their privacy policies to discourage user engagement~\cite{barrett2018critical}.
Some users also mistrust the information presented by companies~\cite{urban2019your}.
\newline 
Low trust in companies appears to be a widespread issue among users. Research shows that, even when users express satisfaction with privacy laws like GDPR aimed at empowering them, they tend to lack confidence in companies' compliance with such laws~\cite{alizadeh2020gdpr,mangini2020empirical,da2018information, pew_cont}.
Consequently, trust may not increase substantially even when companies implement data rights, given the overall low level of trust in companies.
\newline 
Moreover, even if data rights significantly enhance trust in companies, this could have a counter-effect: as users trust service providers more, their willingness to exercise data rights may decrease, as discussed in U2.2~\cite{jilka2021terms, sahqani2021co}.
In summary, the relationship between data rights and trust is complex and warrants further exploration.
\newline 

\indent (C8.3) \textit{Costs}.
While GDPR data rights may yield benefits, they also incur costs for businesses.
As noted in C3 and C5, implementing data rights can be expensive, encompassing employee training and consulting fees.
One study suggests that users perceive GDPR as placing a bureaucratic and infrastructural burden on many otherwise ethical companies due to the misbehavior of others~\cite{alizadeh2020gdpr}.
However, there is also evidence to the contrary.
A survey reveals that most companies do not expect the costs associated with implementing GDPR data rights to significantly impact their business, nor do they foresee a slowdown in their business growth~\cite{company_survey}.
\newline 
Despite these divergent views, we posit that the implementation of GDPR data rights is likely to incur significant costs for most companies.
It is possible that companies underestimate these costs due to a lack of comprehensive understanding of the GDPR.
Although companies claim that costs are manageable, this does not necessarily imply that they can comfortably absorb all expenses related to implementing data rights; they are likely to assign a budget that minimizes the impact on their businesses.
In particular, small companies are constrained to very limited budgets and resources~\cite{sirur2018we}.
As discussed in C3, current investment levels appear inadequate.
Consequently, we believe that the implementation of GDPR data rights is likely to be significantly costly for the majority of companies.

%% file: 006_Regulator.tex
\section{Regulators' perspective}
\label{sec:regulator}

\noindent \textbf{(R1) Are regulators actively monitoring how data rights affect users and companies?}

European regulators adopt a myriad of strategies to gauge and understand the effectiveness of GDPR data rights. Primarily, they interact with citizens through the complaints they receive, serving as a valuable resource for assessing user awareness of data rights~\cite{ico-com, nor-com, kalman2019new, french, cnil-act}. The France Data Protection Authority (DPA) reported a 64\% increase in complaints in 2019 compared to 2018, interpreting this as an indication that EU citizens have ``strongly embraced the GDPR''~\cite{kalman2019new, french}. However, as discussed in U1, this assertion may not hold true. If the volume of complaints was low prior to the enforcement of the GDPR, a 64\% increase may not necessarily imply a substantial rise in user awareness and engagement with GDPR data rights.
\newline 
In addition to public engagement, some DPAs collaborate with corporations to scrutinize digital technologies involved in personal data collection. For instance, the UK DPA partnered with IAB Europe and Google to examine the online advertising ecosystem. Understanding the intricate mechanisms behind online ads is crucial for safeguarding online consumers' data rights~\cite{ico-rtb}.
\newline 
Regulators also recognize the importance of evaluating the GDPR~\cite{reg-eval, edpb}.
Evidence suggests that regulators understand the challenges associated with implementing data rights. Among the most cited difficulties is designing GDPR-compliant ML systems~\cite{g7, ico-ml, por-ml}. Certain regulators have also broached the challenges of implementing GDPR data rights within blockchain systems~\cite{cnil-blockchain}. They actively research ML and blockchain technology and foster discussions through conferences and seminars~\cite{g7,germ-conf, sp-sem, cnil-blockchain}.
\newline 
Moreover, legal experts, alongside regulators, highlight the difficulties in implementing data rights, specifically the right to erasure~\cite{luger2015playing, presthus2021analysis}. In interviews, UK data protection and privacy law experts expressed mixed views on the practicality of the right to erasure~\cite{luger2015playing}.
\newline 
They also express concerns over certain websites deciding not to collect personal information from their users post-GDPR, arguing that it degrades the online user experience and stems from a misunderstanding of the GDPR~\cite{presthus2021analysis}. However, in spite of the challenges and unintended consequences, regulators and legal professionals largely agree that the GDPR has strengthened individual empowerment~\cite{luger2015playing, edpb}.
\newline 

\noindent \textbf{(R2) Are regulators responding and adjusting to the challenges faced by users and companies?}
\label{subsec:effort_r}

Post-GDPR, regulators have enacted supplementary measures to enhance users' data rights. For instance, their websites feature numerous articles and posters promoting the concept of GDPR data rights to users and companies~\cite{cnil,ico,bfdi,cnpd,aepd}.
\newline 
Furthermore, they offer additional guidelines tailored to specific contexts to aid service providers grappling with the integration of GDPR data rights into their systems. The UK regulators, for example, have proposed a guideline for ML systems~\cite{ico-ml2}. This guideline addresses the uncertainties surrounding the right to erasure discussed in C5. The guidance clarifies that when a request to erase personal data is received, ML models built upon the respective data do not need to be deleted if they neither contain nor can be used to infer the data. However, if these conditions are not met, deleting and subsequently retraining the ML models are necessary, which could impose significant costs on the company.
\newline 
Furthermore, the France DPA has offered recommendations for blockchain systems~\cite{cnil-blockchain}. They explicitly advise against registering personal data in clear-text on a blockchain due to the technical impossibility of applying the right to rectification and erasure on such a platform~\cite{cnil-blockchain}.
\newline 
Beyond these, regulators have proposed new guidelines and devised strategies to address various issues relating to online marketing, dark patterns, and cookies~\cite{germ-dark, ico-rtb, cnil-act}. As a result, regulators have been proactive in formulating new guidelines and strategizing to enhance user data rights. However, these guidelines also underscore the substantial costs companies may incur in complying with the GDPR.
\newline

\noindent \textbf{(R3) Do regulators enforce the data rights effectively?}

Ensuring robust enforcement is paramount to the protection of data rights, encompassing the detection and penalization of corporations infringing upon these rights via fines and other sanctions. Despite substantial endeavors by regulators to mitigate the challenges encountered by users and companies (refer to R1 and R2), ample evidence underscores the current inadequacy and deficiency of GDPR data rights enforcement.~\cite{Ireland-paralysis2, sivan2022varieties, ruohonen2022gdpr, barrett2020emerging, wolff2021early, daigle2020eu, akhlaghpour2021learning, custers2018comparison}.

A multitude of studies examining court cases have exposed trends in GDPR enforcement~\cite{ruohonen2022gdpr, barrett2020emerging, akhlaghpour2021learning, wolff2021early,daigle2020eu}. These analyses reveal that imposed fines have typically been marginal in relation to the comprehensive economic and societal repercussions of GDPR noncompliance. Interestingly, even DPAs have adopted a conservative stance, refraining from levying fines to their maximum potential.
\newline
Additionally, an overwhelmingly lesser number of fines were meted out for violations of GDPR articles pertaining to data rights, particularly those besides the rights to information and access. Notably, there were no fine cases relating to the right to eschew automated decision-making, suggesting that EU DPAs face challenges in identifying or adjudicating transgressions of data rights.
\newline
Another key observation is the considerable disparity in both the number of court cases and the quantity of fines levied across Europe, reflecting an inconsistent interpretation and implementation of GDPR data rights articles by EU DPAs. Certain nations, such as the UK, France, and Germany, have demonstrated higher activity in enforcing fines, while others, including Ireland and Slovakia, have displayed relative leniency.

The uneven enforcement across Europe can significantly prolong adjudication in cases that involve multiple countries. A notable report~\cite{Ireland-paralysis2} suggests that Europe has faced hurdles in regulating large tech corporations due to the inefficient operations of Ireland's Data Protection Commission, where many such companies maintain their headquarters. Specifically, an astounding 98\% of GDPR cases linked to Ireland are yet to be resolved. However, EU regulatory authorities have refuted this claim and launched an official investigation~\cite{Ireland-paralysis, Ireland-stats}. The investigation concluded that the practices were appropriate, yet a series of improvements were suggested.

Despite the infancy of enforcement measures, there are promising indicators of maturity. For instance, fines were levied five times more frequently between May 2019 and March 2020 compared to the inaugural year of GDPR implementation~\cite{wolff2021early}. Moreover, DPAs possess the capacity to suspend business operations upon detection of GDPR noncompliance, an authority that could enhance enforcement effectiveness in conjunction with fines~\cite{akhlaghpour2021learning}. Additionally, a survey of 18 DPAs revealed a substantial inclination towards stringent monitoring and sanctioning of noncompliance~\cite{sivan2022varieties}.

Simultaneously, however, several studies indicate that DPAs are grappling with considerable resource constraints, including deficiencies in technical specialists and funding~\cite{Ireland-paralysis2,sivan2022varieties, custers2018comparison}. These limitations could potentially impede the pace of enforcement advancement. Consequently, it is plausible that DPAs heavily depend on user reports to detect data rights infringements~\cite{ico-com, nor-com, kalman2019new, french, cnil-act}.

%% file: 007_Future.tex
\section{Towards a better data rights regime}
\label{sec:future}

Our review indicates that there is substantial room for enhancing data rights implementation. Although our findings focus on the GDPR, they may resonate with other similar laws, such as CCPA. Over 140 countries have comprehensive privacy laws, often drawing from the GDPR or the EU Data Protection Directive~\cite{solove2023rights}. Thus, our study's implications might be applicable to various rights-based laws, raising questions about their effectiveness in truly empowering citizens. For instance, when we examine data consent, which was not within the scope of our analysis, we see that it is also currently ineffective due to reasons similar to those presented in our paper, such as users' non-exercise of consent, inadequate implementation, and the prevalence of dark patterns~\cite{kampanos2021accept, kulyk2020has, mehrnezhad2020cross, hu2019characterising, sanchez2019cookie, santos2021cookie, bauer2021you, strycharz2021no, gray2021dark, darkpatters}.
\newline 
Nevertheless, our findings also suggest the potential for a more robust data rights framework. Based on our study's insights from the 15 key questions, we propose recommendations for both policymakers and Computer Science (CS) communities, without drastically altering the existing data rights paradigm. Additionally, we contemplate the possibility of a new data rights framework leveraging privacy-enhancing technologies. After discussing the potential of these technologies to enhance data rights, we conclude by outlining unresolved tensions in the current system. Fig.~\ref{fig:recommendation} in Appendix shows how recommendations and tensions are derived from the analysis of the 15 questions.

\subsection{Recommendations for Regulators \& CS Communities}

\hspace{4mm}\textbf{Education and training}.
In light of the lack of awareness about data rights among users and companies (see U1 and C1), policymakers should consider institutional mechanisms and processes for formal education and training given the education effectiveness (see U1.4). This can help bridge the knowledge gap in data rights, particularly among less internet-savvy users (see U1.2). In addition, companies should be encouraged to conduct regular employee training sessions beyond the customary one-time training (see C3). Such measures can significantly enhance users' and companies' knowledge and awareness of data rights.

\indent \textbf{Standardization}. The current GDPR without standardization has led companies to provide data rights to users in different ways using different interfaces and modes of communication. 
For example, if users want to exercise their data rights, they should email some companies and call other companies. 
Some companies even use social media to receive data rights requests from users~\cite{aydin2020gdpr}.
Moreover, some data rights interfaces can intentionally employ dark patterns, rendering navigation confusing (see C6.1). This lack of standardization can inhibit users from effectively exercising and even being aware of their data rights (see U1.3), thus providing companies an avenue to dodge data rights obligations. 

Standardization can also be needed in the enforcement of GDPR data rights. Currently, regulatory authorities face challenges due to inconsistent enforcement practices across Europe (refer to R3). The introduction of standardization could foster a more robust and efficient data rights regime.

\indent \textbf{Assessing implementation costs}.
The accurate estimation of user empowerment costs is vital for pragmatic policy development. Implementing data rights poses challenges and cost burdens for companies (see C5 and C8.3).  Current guidelines offer some support to companies grappling with data rights implementation. However, the practicality of these guidelines is debatable due to the substantial costs incurred to ensure compliance (see R2).
In some instances, regulators acknowledge these high costs. For instance, in a supplementary guideline for ML companies, regulators suggest that having well-organized systems might reduce costs~\cite{ico-ml2}. Yet, companies may argue that even with streamlined systems, the cost of retraining and redeploying ML models remains high.

\indent \textbf{Strict enforcement}.
Stricter regulatory monitoring and law enforcement are crucial. The current implementation of data rights by almost all companies is defective (see C4). Moreover, the current enforcement is deficient along with a lack of resources like technical expertise and funding (see R3). The user perceptions that companies would not comply with GDPR data rights can discourage individuals from exercising their data rights (see C8.2), and some companies may exploit perceived lax oversight to evade data rights obligations (see C2.1). To resolve these issues, enhanced enforcement would be essential. Regulators can leverage research findings to prioritize enforcement efforts for specific applications, given the variation in compliance rates (see C4.1).

\indent \textbf{Automated tools for assisting in data rights implementation and enforcement}.
The enforcement of laws poses significant technological challenges to companies and regulators (see C5, C7, and R3). Coupled with companies' struggles to implement data rights and assess compliance status, these challenges underscore the need for automated compliance tools. Such tools could ease data rights implementation for companies and aid regulators in compliance monitoring and law enforcement.
Consider, for instance, an automatic tool that evaluates a company's compliance without manual intervention. This tool could help companies identify their compliance status and subsequently enhance data rights implementation. Also, some automatic tools can reduce the complexity of implementing a data right. Many researchers have pursued the development of such tools~\cite{wang2022privguard,fernandez2018special,azraoui2014ppl,chowdhury2014temporal,maniatis2011you,mansour2016demonstration,saroiu2015policy,sen2014bootstrapping,trabelsi2011ppl,wang2019riverbed, voloch2021implementing, li2018coconut, li2021honeysuckle, gardner2022helping}, but efforts are disproportionally focused on the right to information. Despite significant progress, the real-world effectiveness of these tools remains questionable (see C4 and C5), suggesting that further work is needed to improve the data rights regime.

\subsection{Privacy-Enhancing Technologies}

The implementation of data rights presents numerous technical challenges. Furthermore, the unique nature of data amplifies these challenges, making it difficult for data owners to assert clear data rights. The current rights-based privacy approach often simplifies data as a property in the modern economy, which may not fully capture its characteristics. Privacy-Enhancing technologies can be instrumental in addressing these issues~\cite{ GARRIDO2022103465, le2016whom}.
\newline
Data differ from physical objects in many ways. For example, data are non-rivalrous, meaning that once shared, data cannot be easily retracted, posing challenges in controlling its usage. Both users and companies struggle with this issue, often lacking a full understanding of complex data practices. Additionally, data dependencies and externalities exist where processed data can leak information about the original data, and data related to one entity can reveal information about others. For instance, social media data shared by a user often include information about their social network. Therefore, a user's exercise of data rights can inadvertently impact others.
\newline
Consequently, we cannot apply property governance principles to data without innovation. Technological advances are needed to enforce data rights, as traditional solutions prove insufficient. For example, traditional data encryption cannot protect data in use~\cite{datainuse}, and anonymization alone cannot prevent leakage of sensitive identity-related information through inference. The ideal technology should fulfill two key requirements: (1) controlling data usage to enable computation without sharing the original data, and (2) preventing sensitive information from being leaked through the processed output.
Fortunately, several emerging technologies possess properties that, in theory, meet these criteria. Secure computing techniques, such as trusted hardware, multi-party computation, zero-knowledge proof, and fully homomorphic encryption, enable data processing without revealing raw data. Differential privacy ensures that computation output does not leak individual-specific information. Federated learning facilitates distributed ML model training while keeping data local. Additionally, a distributed ledger provides immutability, serving as an effective auditing tool to ensure compliance in data usage. 
\newline
Through the adoption of these advanced technologies, companies can utilize and process extensive personal data without requiring users to share excessive information, addressing many privacy concerns and technical challenges of data rights. This new data rights regime could offer users greater control and relieve regulatory bodies of monitoring burdens. 
\newline
While these technologies have ideal properties to process sensitive data, they are still maturing and have not yet fulfilled their promise~\cite{277172, GARRIDO2022103465, le2016whom}.
In addition, the new data rights regime is not a panacea. Digital divide, psychological, and economic factors may still prevent users from fully exercising their data rights (see Section~\ref{sec:user}). Despite this, we posit that the adoption of these technologies could enhance the effectiveness of the rights-based approach.

\subsection{Persistent Challenges in Rights-Based Privacy Regimes}

Despite the advancements in the proposed recommendations and privacy-enhancing technologies, persistent tensions present considerable challenges within rights-based privacy regimes.
\newline
\indent \textbf{Data rights vs. burdens}.
The empowerment of users can inadvertently create burdens, a key reason many users refrain from exercising their data rights (see U2.2, U3.1, and U3.2). Although high usability data rights tools are anticipated to alleviate this issue, their real-world efficacy remains uncertain (see C6.3.2). Finding a definitive solution to this problem and motivating users to actively exercise their data rights are still open challenges.
\newline
\indent \textbf{Users vs. companies vs. regulators}.
Conflicting interests among users, companies, and regulators, coupled with the uncertainty regarding the alignment of data rights with these interests (see U4 and C8), add further complexity to the situation. While regulators prioritize user privacy, companies aim to expand their business operations. User interests vary, with some prioritizing privacy and others valuing service quality or time efficiency. Users' decisions to share personal information with companies depend on these diverse interests. These divergent values among the stakeholders make the attainment of effective data rights a complex endeavor.

%% file: 008_Alternative.tex
\section{Alternative Approaches to Data Rights}

Throughout this paper, we have identified that one of the primary issues with rights-based approaches is the reliance on questionable user assumptions. Few users are willing to undertake the burden required to exercise their data rights. Consequently, an increasing number of scholars question the rights-based approach for this reason~\cite {solove2023rights, hartzog2022legislating, khan2019skeptical, alternative, benthall2021data, regan1995legislating, viljoen2021relational, huq2021public, gordon2022case, fuchs2011towards}. Many have proposed alternative solutions that supplement rights allocations, often by altering the relationship dynamics between data subjects (i.e., users) and controllers (i.e., companies).
\newline 
Responsibility-based approaches shift more privacy obligations to data controllers~\cite{solove2023rights}. A responsibility regime could require data controllers to assume the burden of making decisions with personal information as a cost. For example, a policy might mandate controllers to review and correct 5\% of their files annually, as opposed to a rights-based regime where data subjects must \textit{initiate} the laborious process of accessing, reviewing, and requesting corrections.

Some commentators favor the creation of an explicit duty of loyalty on data controllers towards data subjects, which is called ``data fiduciary'' or ``trust''~\cite{hartzog2022legislating,balkin2015information, balkin2020fiduciary,haupt2020platforms,edwards2004problem}. 
This duty of loyalty would obligate controllers to design services and operate in ways that do not conflict with users' interests. Duties represent potent obligations; if imposed within the information economy, they would challenge practices like ``nudging,'' wherein data is used to manipulate or unfairly target users. Loyalty obligations cannot be easily undone like rights-based approaches, as courts tend to critically construe waivers of duties against the company owing the obligation. Consequently, fine print, dark patterns, and manipulative interfaces would fail to undermine duties~\cite{hartzog2022legislating}. However, duties do not offer a definitive solution for data power issues, as data controllers, unlike doctors and lawyers, must act in their economic self-interest. A duty of loyalty to data subjects is at odds with the longstanding consensus that companies' primary moral imperative is to maximize shareholder value~\cite{khan2019skeptical}.

Rights-based approaches position individuals as pivotal actors in privacy power dynamics. Some commentators advocate for overturning this user-centricity, suggesting a collective-societal approach where ``group privacy'' becomes the central principle~\cite{regan1995legislating,viljoen2021relational,huq2021public, gordon2022case, fuchs2011towards}. Priscilla Regan was an early advocate for this collective viewpoint, foreseeing the power shifts resulting from data proliferation and arguing that collective privacy interests would better serve the political landscape~\cite{regan1995legislating}.

The recent U.S. Supreme Court decision in \textit{Dobbs}, which compromises women's access to reproductive health, serves as a poignant illustration of the importance of collective privacy interests. This ruling marks a pivotal moment in privacy policy discourse, highlighting the sweeping societal ramifications of such decisions. The post-\textit{Dobbs} era ushers in a heightened data privacy peril for all U.S. women. Unintentional disclosure through commercial advertising, whether by purchase history or search patterns, could potentially expose their non-compliance with state anti-abortion laws. This stark reality underscores the pressing need for fortified collective privacy safeguards.

Collective privacy regimes could assume various forms. For example, some propose that governments establish ``public trusts'' for their residents' personal data. This would treat personal information in public records as a publicly-managed asset, rather than a commodity freely available for anyone to download. The state would then have the authority to permit the usage and processing of the data, ensuring it benefits the wider public rather than a select few companies~\cite{huq2021public}. Another proposal is the establishment of a ``collective perspective'' on data held and processed by digital platforms~\cite{gordon2022case}. This perspective would grant a third party (e.g., a regulator or a consumer collective organization) ongoing insight into the collected and processed data, as well as its correlation with personalized content. The aim is to enable such third parties to comprehend, identify, quantify, and address harms driven by personalization~\cite{gordon2022case}.
\newline 
In our view, society is currently experiencing a shift akin to the industrial revolution. Just as industrialization introduced new challenges related to labor, safety, power relations, and the environment, the information economy is generating previously inconceivable conflicts. The solutions to these tensions are unlikely to be apparent or easily executed. Rights-based approaches represent the ``first draft'' in privacy history, and these might be complemented or replaced by responsibility regimes, a trust approach, or a reorientation towards collective privacy concepts.

%% file: AI_privacy.tex
\section{Emerging Privacy Challenges in the Age of AI}
\label{sec:ai}

The rapid proliferation of AI systems and the escalating race to train ever-larger models have surfaced many privacy concerns~\cite{miller2024privacy}. Chief among them is the fact that training corpora---often scraped indiscriminately from the open web---routinely contain private or sensitive information. A growing body of work demonstrates that large models can memorize such data and later reveal it at inference time~\cite{xummdt,aditya2024evaluating,jegorova2022survey}. Consequently, the composition of training datasets and the policies that govern their use matter profoundly.

Today, AI developers adopt heterogeneous privacy policies. OpenAI, for instance, retains user-model conversations for model improvement or training unless users explicitly opt out~\cite{openai_privacy}. Anthropic, by contrast, pledges not to incorporate user data into training by default~\cite{anthropic_privacy}. The absence of a harmonized industry standard and a comprehensive legal framework for training data governance exacerbates user data rights~\cite {trainingdata_legal}. As models continue to scale, the need for statutory clarity intensifies.

Legal safeguards, however, are merely one piece of the puzzle. If individuals wish to invoke the ``right to erasure,'' technical mechanisms must exist to remove their data from a trained model. Several approaches have been explored. Data-centric techniques attempt to delete sensitive text before training~\cite{kandpal2022deduplicating,lison2021anonymisation}, while others rely on differential privacy to provably limit memorization~\cite{dwork2008differential,anil2021large,yu2021differentially}. Both strategies generally require retraining the model, incurring prohibitive computational costs at frontier scales.
To avoid full retraining, researchers have proposed \emph{knowledge unlearning}, which updates only a small subset of parameters to excise targeted knowledge~\cite{jang2022knowledge}. Although promising, current methods remain in their infancy and can trigger catastrophic forgetting or other safety failures~\cite{si2023knowledge}. Further work---both technical and legal---is therefore essential.

Privacy risks are not limited to memorization. Even without data leakage, large multi-modal models can infer personally identifiable information from inputs~\cite{xummdt}. As capabilities grow, so too do privacy harms, underscoring the urgency of holistic solutions that span policy, model architecture, and deployment practices. 

%% file: 009_Conclusion.tex
\section{Conclusion}
\label{sec:conclusion}

Privacy is a critical concern in the information economy, yet it is often compromised. In response, data rights have emerged as a novel facet of human rights. However, as our comprehensive analysis of existing literature reveals, the implementation and acceptance of data rights are presently inconsistent, even with individuals' willingness to exercise data rights varying significantly. While some empirical studies demonstrate the efficacy of data rights in specific contexts, our findings lend credence to those skeptical of the general efficiency of current rights-based systems.

Our evaluation of data rights illuminates the areas that need improvement in order to establish a robust data rights regime. Based on these findings, we propose recommendations for an improved rights-based regime and discuss alternative approaches aimed at restructuring the power dynamics between users and companies. We highlight the complexities of building an effective privacy protection system, leaving many open questions for researchers, developers, and regulators.

Through our new data rights assessment framework, we aim to provide a comprehensive understanding of the current status of data rights and offer insights to guide future research and policy-making in enhancing privacy interests.

%% file: 100_Appendix.tex
\section{Framework Questions}
\label{app_sec:Framework_questions}

\subsection{Users}
\label{subsec:Users_questions}

Data rights primarily aim to benefit individuals. We identify four key questions concerning users:
\smallskip

\noindent\textbf{(U1) User knowledge: Are users informed about and understand data rights?} 
It is critical that users know these rights and how to exercise them.
 
\noindent\textbf{(U2) User perception: How do users perceive data rights?} 
User perception of rights may affect users' willingness to exercise them.

\noindent\textbf{(U3) User action: Do users exercise data rights in practice?}
Rights' effectiveness depends on whether users exercise them.

\noindent\textbf{(U4) Effect on users: Do data rights benefit users?} 
There might be a gap between the expected and actual benefits of data rights.

\subsection{Companies \& Developers}
\label{subsec:companies_questions}

Companies play a crucial role in operationalizing these rights. We identify the following key questions:
\smallskip 

\noindent\textbf{(C1) Company knowledge: Are companies informed about and understand data rights?} 
Facilitating data rights correctly can only happen when companies understand and align with them.

\noindent\textbf{(C2) Attitude towards compliance: What is companies' attitude toward complying with data rights?}
Undervaluing the implementation of data rights will undermine their effectiveness.

\noindent\textbf{(C3) Administrative efforts: Have companies made internal changes to support data rights?}
Complying with the law may entail significant changes in a company's internal policies, processes, and IT infrastructure, etc., and requires evidence of such efforts.

\noindent\textbf{(C4) Compliance with the law: Do services operated by companies enable users to exercise data rights?}
It is critical to inquire whether companies lawfully facilitate users' data rights.

\noindent\textbf{(C5) Significant difficulties: Is it technically and managerially feasible for companies to implement data rights?}
It is essential to explore whether there are difficulties in implementing data rights. 


\noindent\textbf{(C6) Usable design: Does the user process of exercising data rights have high usability?}
A burdensome process of exercising rights would hinder users from exercising their rights.

\noindent\textbf{(C7) New threats or flaws: Are data rights being implemented without creating new technical flaws or threats?}
We also investigate whether data rights create negative side effects.

\noindent\textbf{(C8) Effect on companies: Do data rights produce positive effects for complying companies?}
It is worth investigating whether companies complying with data rights can also benefit in some dimension.

\subsection{Regulators}
\label{subsec:regulators_questions}

We discuss the role of legislators and regulators in monitoring and adjusting the efficacy of the regime they devised, and in enforcing the data rights as well:
\smallskip

\noindent\textbf{(R1) Monitoring: Are regulators actively monitoring how data rights affect users and companies?} 
It is critical to research how and whether regulators monitor the data rights effect on users and the industry.

\noindent\textbf{(R2) Response and adjustments: Are regulators responding and adjusting to the challenges faced by users and companies?}
It is important to examine the regulator's efforts to amend the legal text or provide supplementary guidelines in response to possible negative effects/challenges on society.

\noindent\textbf{(R3) Enforcement: Do regulators enforce the data rights effectively?} It is to examine whether the current enforcement is proper to attain effective rights-based regimes.

\begin{table*}[ht]
    \centering
    \caption{Eight GDPR data rights}
    \label{tab:right}
    \begin{tabular}{m{0.31\textwidth}|m{0.7\linewidth}}
     \hline
    \multicolumn{1}{c}{Data Right}& \multicolumn{1}{c}{Description}\\
    \hline
    \hline
    \makecell{Right to information \\(Arts. 13\&14 of the GDPR)} & Service providers that collect users' data must provide data subjects with information about when and why they collect data, the way to exercise their data rights in the service, etc. \\
    \hline
    \makecell{Right to access \\(Art. 15 of the GDPR)} & The data subject should be able to access their personal data that companies process and ask for information on companies' data processing.\\
    \hline
    \makecell{Right to rectification \\(Art. 16 of the GDPR)} & Subjects should be able to change and update their personal data that service providers store.  \\
    \hline
    \makecell{Right to erasure \\(Art. 17 of the GDPR)} & Data subjects can erase their personal data by sending a request to the company under certain conditions (e.g., in the case where they withdraw their consent to data processing). \\
    \hline
    \makecell{Right to restriction of processing \\(Art. 18 of the GDPR)} & Users can ask companies to restrict the processing of their data under certain conditions (e.g., in the case where they contest the accuracy of the data). \\
    \hline
    \makecell{Right to data portability \\(Art. 20 of the GDPR)} & Users should be able to receive the personal data held by companies in a machine-readable form and send the data to other companies. \\
    \hline
    \makecell{Right to object\\(Art. 21 of the GDPR)} & Users can object to processing their personal data under certain conditions. The right to object to direct marketing is absolute.\\
    \hline
    \makecell{Right to avoid automated decision-making \\(Art. 22 of the GDPR)} & 
    As the last data right of the GDPR, this right allows users to request not to be subject to a decision based solely on automatic processing, including profiling. 
    \end{tabular}
\end{table*}

\begin{table*}[ht]
    \centering
    \caption{Codebook used in the analysis of our dataset}
    \label{tab:code}
    \begin{tabular}{c|m{0.6\linewidth}|c}
    \hline
        Label &\hfil Description \hfill& Relevant ones of 14 factors \\
         \hline
         \hline
         Policy-comp & Investigate whether privacy policies are written in compliance with GDPR & C4 \\
         \hline
         Policy-read & Investigate how readable privacy policies are & C6\\
         \hline
        NewDesign & Propose a new system design to better support data rights & C6\\
        \hline
        Cookie-comp & Investigate whether cookie notices comply with GDPR & C4\\
        \hline
        Cookie-read & Investigate how readable cookie notices are & C6\\
        \hline
        User-know & User knowledge or awareness of data rights & U1\\
        \hline
        User-th & Users' general thoughts on data rights (e.g., Perception, expectation, and willingness)  & U2\\
        \hline
         User-behave & User behaviour and experience regarding data rights; in particular, how they handle their data in the real world & U3\\
        \hline
        User-effect & Data rights' effects on users (e.g., their trust in companies, their privacy concerns, behaviour of sharing their personal information) & U4\\
        \hline
        Request-comp & Investigate whether service providers comply with GDPR by sending a data rights request to them & C4\\
        \hline
        System-comp & Investigate whether service providers comply with GDPR through a system analysis & C4\\
        \hline
        RWdata-comp & Investigate whether service providers comply with GDPR through an analysis of real-world data such as social media data and customers' reviews & C4\\
        \hline
        Usab & Investigate usability of exercising data rights (except for an analysis of privacy policies) & C6\\
        \hline
        Dev-know & Developers' (or companies') knowledge or awareness of data rights & C1\\
        \hline
        Dev-th & Developers' (or companies') general thoughts on data rights (e.g., Perception, willingness, and attitude) & C2, C5\\
        \hline
        Dev-effort & Developers' (or companies') efforts and experience regarding the implementation of data rights; & C3, C4\\
        \hline
        Dev-effect & Data rights' effects on companies (e.g., (dis)advantages of GDPR) & C8\\
        \hline
        Reg-effort & Regulator' efforts to support data rights & R1, R2, R3\\
        \hline
        Law-th & Perception and thoughts of law experts & R1, R2\\
        \hline
        Infra & Investigate companies' infrastructures that have changed to support data rights & C3\\
        \hline
        Dark & Dark patterns & C6\\
        \hline
        Flaw & Investigate whether there are flaws in the current implementation of data rights & C7\\
        Fines & Investigate whether fines are sufficient to enforce the data rights & R3\\   
    \end{tabular}
\end{table*}

\onecolumn
\clearpage
\begin{longtable}{>{\centering\arraybackslash}m{0.08\textwidth}|>{\centering\arraybackslash}m{0.15\textwidth}|m{0.74\textwidth}}
\caption{Overview of key findings from GDPR data rights evaluation}\label{tab:framework}\\
\hline
\textbf{Agents} &\multicolumn{1}{c|}{\textbf{Question}} &\multicolumn{1}{c}{\textbf{Key Findings}}\\
\hline\hline
\multirow{11}{*}{}& \multirow{4}{*}{} & \textbf{(U1.1)} User knowledge varies across data rights. \\
&\textbf{U1}: User knowledge& \textbf{(U1.2)} The digital divide leads to a gap in GDPR data rights knowledge. \\
&&\textbf{(U1.3)} The level of user knowledge of data rights depends on context.\\
&&\textbf{(U1.4)} Education can foster user knowledge of data rights.\\
\cline{2-3}
&\multirow{2}{*}{\vspace{-8mm}\textbf{U2}: User perception}& \textbf{(U2.1)}  Many users express interest in having data rights across various contexts. \\
\textbf{Users}&& \textbf{(U2.2)} However, many others don't also feel the necessity of data rights due to high trust in companies that collect their data, indifference towards using their data, privacy fatigue caused, fear from learning about personal data processing, and a lack of understanding data rights.\\
\cline{2-3}
& \multirow{3}{*}{} & \textbf{(U3.1)} The interest in having data rights doesn't directly lead to the actual exercise of rights by users. \\
&\textbf{U3}: User action&\textbf{(U3.2)} The evidence suggests that users don't exercise their data rights very often in general. \\ 
&&\textbf{(U3.3)} But exercise can become active when the information is especially sensitive.\\
\cline{2-3}
& \multirow{2}{*}{\textbf{U4}: Effect on users} & \textbf{(U4.1)} Per the potential benefits reviewed, although having data rights can reduce users' perceived privacy concerns in general, how data rights are implemented in practice has a significant influence as well. Moreover, the right to avoid automated decision-making little improves user perception. \\
&& \textbf{(U4.2)} It is also unclear whether ``GDPR'' data rights encourage user privacy-seeking behavior; how to implement data rights affects whether the rights would encourage users' privacy-seeking behavior.\\
\hline
\hline
\multirow{18}{*}{}& \textbf{C1}: Company knowledge & Many companies aren't fully aware of GDPR data rights and don't understand them well. \\ 
\cline{2-3}
& \multirow{2}{*}{\makecell{\textbf{C2}: Attitude \\towards compliance}} &  \textbf{(C2.1)} 
Many developers remain unaware, indifferent, or skeptical about requirements to comply with the obligation to provide data rights. \\
&&\textbf{(C2.2)} However, many other companies also claim to implement rights according to the GDPR or say it is important.\\
\cline{2-3}
& \textbf{C3}: Administrative efforts & Many companies have taken internal measures to facilitate their GDPR compliance, including employee training, changes in management, and procuring professional advice. However, some companies have decided to evade certain GDPR data rights. \\
\cline{2-3}
& \multirow{2}{*}{\makecell{\textbf{C4}: Compliance \\with the law}} & \textbf{(C4.1)} Many companies do not fully provide the right to information, and the level of compliance varies significantly across service applications. \\
& & \textbf{(C4.2)} The finding (C4.1) also applies to various data rights: the rights to access, data portability, and erasure. Additionally, many companies may not even be aware of their GDPR compliance status.\\
\cline{2-3}
&  \multirow{6}{*}{} &Companies complain about several technical and managerial challenges in implementing data rights.\\
&  &\textbf{(C5.1)} The right to erasure is considered particularly challenging amongst companies due to the process complexity and the incompatibility between the requirements of the right and certain central technologies companies deploy.\\
\textbf{Companies or developers}& \textbf{C5}: Significant difficulties & \textbf{(C5.2)} Implementing the rights to information and access is challenging due to complex data processing by third parties. Moreover, explaining ML algorithms to general users is difficult under the right to information, as these algorithms are often seen as black boxes and too technical for user comprehension. User identification is also considered a challenge in enforcing the right to access. \\
&   & \textbf{(C5.3)} Heterogeneous data structures among companies make the right to data portability difficult to enforce.  \\
&   & \textbf{(C5.4)} The right to avoid automatic decision-making causes several issues such as decision fatigue and scalability from human review. \\
&  & \textbf{(C5.5)} Data rights may have a disproportionate impact on certain companies, given the heterogeneous challenges faced by different organizations. \\
\cline{2-3}
& \multirow{3}{*}{} &\textbf{(C6.1)} Many studies show poor usability of several rights, including the right to information, the right to access, the right to erasure, and the right to object. \\
&\textbf{C6}: Usable design&\textbf{(C6.2)} But, there is conflicting evidence regarding the usability of data rights, as some companies successfully provide data rights with good usability. Therefore, the level of usability depends on the specific application or implementation. \\
& & \textbf{(C6.3)} Contrary to the wishes of many, it is currently unclear whether dedicated tools designed for high usability are effective or lead to increased use of data rights in the real world.\\
\cline{2-3}
& \textbf{C7}: New threats or flaws & The current implementation of the access right increases the risk of information leakage to unauthorized third parties. \\
\cline{2-3}
& \multirow{2}{*}{\makecell{\textbf{C8}: Effect on \\ companies}} & \textbf{(C8.1)} The impact of data rights on compliant companies is unclear.  While GDPR data rights can enhance internal data management, there is a possibility that users may hesitate to share their data under these rights, which hinders efficient data utilization by companies. \\
&&\textbf{(C8.2)} The relationship between data rights and user trust in companies is complex. While implementing data rights can enhance a company's corporate image, the impact may not be substantial if the initial image is already highly negative. \\
&&\textbf{(C8.3)} Complying with the GDPR data rights is anticipated to entail significant costs for companies.\\
\hline
\hline
\multirow{3}{*}{}& \textbf{R1}: Monitoring & Regulators communicate with citizens, research, and hold conferences and seminars to understand the effects of data rights on users and companies. 
They are aware of some of the challenges in implementing data rights, but they also evaluate that the GDPR has overall empowered people.\\
\cline{2-3}
\textbf{Regulators} & \textbf{R2}: Response and adjustments &  Regulators provide additional guidelines to users and companies. However, it remains to be seen whether the proposed guidelines will be usable and achievable.  \\
\cline{2-3}
& \textbf{R3}: Enforcement &  The enforcement of GDPR data rights is currently lacking, with small fines, limited technical expertise, and funding challenges faced by regulatory authorities. Additionally, there is heterogeneity in enforcement across Europe.
\end{longtable}

\begin{figure}[!ht]
    \centering
    \includegraphics[width=0.7\columnwidth]{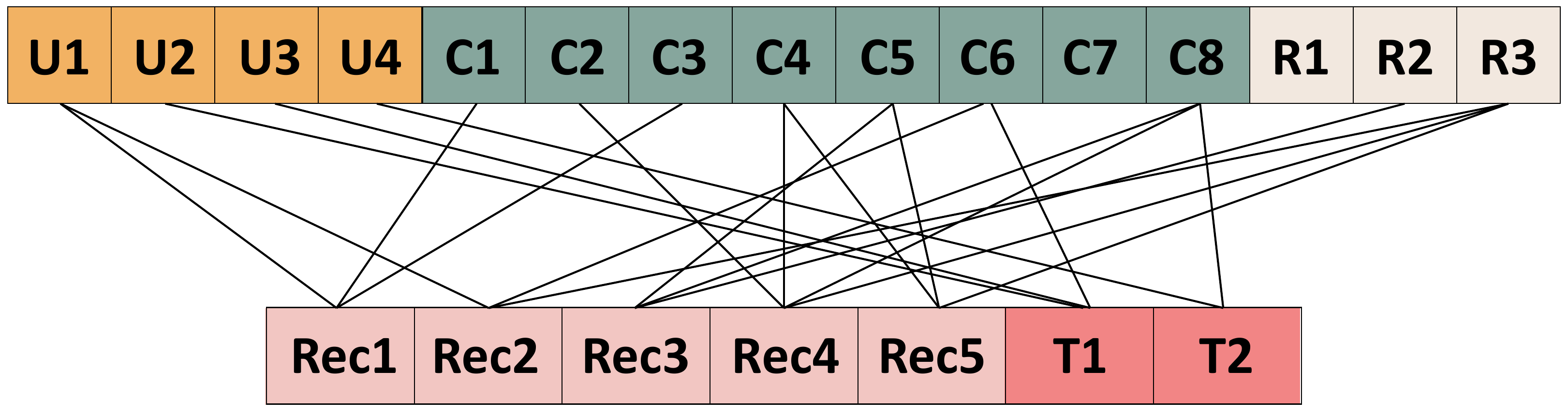}
    \caption{A connection between our data rights analysis and recommendations/tensions presented in Section~\ref{sec:future}. \normalfont{Rec1, 2, 3, 4, and 5 indicate education, standardization, assessing implementation costs, strict enforcement, and automated tools for assisting in data rights implementation/enforcement, respectively. T1 and T2 indicate the two tensions: data rights vs. user burdens, and conflicting interests among users, companies, and regulators.}}
    \label{fig:recommendation}
\end{figure}